\documentstyle[12pt]{article}
\renewcommand{\baselinestretch}{1.5} 
\begin{document} 
\thispagestyle{empty} 
\begin{flushright}
UA/NPPS-1-04\\
\end{flushright}
\begin{center}
{\large{\bf
A numerical algorithm for computing the complete set
of one-gluon loop diagrams in QCD
on the basis of a single master expression
\\}}
\vspace{2cm} 
{\large A. S. Kapoyannis*, A. I. Karanikas and C. N. Ktorides}\\ 
\smallskip 
{\it University of Athens, Physics Department\\
Nuclear \& Particle Physics Section\\
Panepistimiopolis, Ilisia GR 157--71, Athens, Greece}\\
\vspace{1cm}

\end{center}
\vspace{0.5cm}
\begin{abstract}

A numerical program is presented which facilitates the computation of
the full set of one-gluon loop diagrams (including ghost loop
contributions), with M attached external gluon lines in all possible
ways. The feasibility of such a task rests on a suitably defined master
formula, which is expressed in terms of a set of Grassmann and a set of
Feynman parameters, the number of which increases with M. An important
component of the numerical program is an algorithm for computing
multi-Grassmann variable integrals. The cases M=2, 3, 4, which are the
only ones having divergent terms, are fully worked out. A complete
agreement with known, analytic results pertaining to the divergent terms
is attained.
\end{abstract}

\vspace{3cm}


*Corresponding author.
{\it E-mail address:} akapog@nikias.cc.uoa.gr (A. S. Kapoyannis)

\newpage
\setcounter{page}{1}

{\large \bf PROGRAM SUMMARY}

{\it Title of program: DILOG}

{\it Program obtainable from: CPC Program Library, Queen's University of
Belfast, N. Ireland}

{\it Computer for which the program is designed on and others on which it
has been tested: Personal Computer}

{\it Operating systems or monitors under which the program has been tested:
Windows 98}

{\it Programming language used: FORTRAN 77}

{\it Memory required to execute with typical data: 73 728 words}

{\it No. of bits in a word: 64}

{\it No. of processors used: one}

{\it Has the code been vectorized or parallelized?: no}

{\it No of bytes in distributed program, including test data, etc: 589 824}


{\it Nature of Physical problem}\\
The computation of loop diagrams in QCD with many external gluon lines is a
time consuming task, practically beyond reasonable reach of analytic
procedures. In this paper we apply recently proposed master formulas for
the computation of one-loop gluon diagrams in QCD with ah arbitrary number,
$M$, external gluon lines.

{\it Method of solution}\\
The Grassmann variables in the master expressions and their properties are
represented in suitably defined integer matrices, in order to carry out the
Grassmnan integration first. The parametric functions are handled by an
integer representation. The output of the program is the complete analytic
result (tested for validity up to $M=4$).

{\it Restrictions on the complexity of the problem: $M$ must one exceed
4, though extension to higher values is simple.}

{\it Typical running time: 25 seconds for $M=4$}

\newpage

{\large \bf LONG WRITE-UP}

{\large \bf 1. Introduction}

Quantum Chromodynamics (QCD) enjoys, in our days, universal acceptance as the 
fundamental theory for the strong interaction. As a quantum field theoretical system, 
QCD has been extensively applied to situations in which its perturbative content provides 
a dependable computational tool. It is, in fact, within the framework of this perturbative 
content that QCD has successfully confronted the quantitative description of the 
multitude of scattering processes, which probe strong interaction dynamics at high 
energies. Admittedly, the study of the non-perturbative domain of the theory offers 
intriguing challenges. Nevertheless, the immediate need to confront recent measurements 
coming from the HERA and Tevatron particle accelerators as well as the expected ones, 
in the near future, from the LHC accelerator continues to put perturbative QCD (pQCD) 
to the forefront of theoretical activity\footnote{Non-perturbative input in the relevant 
computations enters in the form of `initial' phenomenological information, with pQCD 
taking charge from thereon.}.

Given the non-abelian structure of QCD, the (by far) most demanding component of 
the theory in relation to perturbative calculations is its gluonic, as opposed to its quark, 
sector\footnote{For that matter, this is more so the case for the non-perturbative domain 
of the theory.}. In particular, perturbative computations involving Feynman diagrams 
with gluon/ghost loops become, to say the least, quite monstrous. During the last decade 
or so various methods, aiming to expedite Feynman diagram computations in QCD, have 
been proposed whose basic feature is that they rely in a first, rather than the usual second, 
quantization approach to the formulation of the theory. Corresponding attempts have 
employed either strings [1-3], or world-line paths [4-10] as their underlying basic agents. 
Within the framework of the latter case, two of the present authors, Refs [9,10],
were involved in work which led to the formulation of a set of master expressions,
that condense the multitude of all Feynman diagrams entering
a given configuration determined by the number of loops and the number of external 
propagators attached on them. To be more precise, the derived expressions  go up to two 
loop configurations, nevertheless the `logic' of the construction can be extended to loops 
of higher order. Even so, the analytical confrontation of a two loop situation with four 
`external' gluon lines constitutes a challenging enough problem [11].

The basic feature of the master expressions arrived at in Refs [9,10] is that they are 
furnished in terms of a set of Grassman and a set of Feynman variables. Once integrations 
over these two sets of variables are performed one obtains the full result, i.e. the one 
which, for the given configuration, contains the contribution of all Feynman diagrams at 
once. It is obvious, even before laying an eye on these master formulas, that in order to 
put them into practical use, one should employ suitable computational methods for 
confronting them. It is the aim of this paper to present such a program, which will be 
applied to the one gluon loop case for two, three and four external gluonic lines. Given 
that one part of our program deals with the confrontation of multi-Grassmann variable 
integrals, which are entangled with expressions involving additional variables (also in 
line to be integrated over), it is hoped that it could find applicability to other situations, 
where Grassmann variables also make their entrance. 

Our paper is organized as follows. In the following section we present the battery of 
formulas, which are associated with the master expression corresponding to one 
gluon/ghost loop with $M$ external gluon attachments in all possible ways. We intend to 
consider the cases $M=2$, 3 and 4, which exhibit divergent terms\footnote{As
expected, the aforementioned master expressions implicate the absence of
divergent terms for $M>4$, cf. Ref [9].}, in addition to finite ones.
In section 3 we describe the structure of the program, while section 4 presents our results, 
with the first two cases being explicitly displayed. Finally, our concluding
remarks are made in section 6.

\vspace{0.3cm}
{\large \bf 2. The one loop master formula}

Consider a configuration consisting of one gluon/ghost loop onto which $M$ external 
gluon lines, with corresponding momenta $p_1\cdot\cdot\cdot p_M$ are attached (see 
Figure). According to Ref [9], the master expression, which summarizes the total 
contribution from all Feynman diagrams pertaining to this configuration is given by
\[
\Gamma_1^{(M)}(p_1,\dots,p_M)=-\frac{1}{2} g^M (2\pi)^4
\delta^{(4)} \left(\sum_{n=1}^{M}p_n\right)
Tr_C (t_G^{\alpha_M} \cdots t_G^{\alpha_1}) \frac{1}{(4\pi)^2}
\int_0^\infty dT T^{M-3} \times
\hspace{10cm}\]
\[
\times \left[ \prod_{n=M}^1 \int_0^1 du_n \right]
\theta (u_M,\ldots,u_1) F^{(M)}(u_1,\ldots,u_m;T)
\exp\left[T\sum_{n<m} p_n \cdot p_m G(u_n,u_m) \right]
\hspace{10cm}\]
\begin{equation}
+ permutations \hspace{10cm}\;,
\end{equation}
where $g$ is the coupling constant of the theory, the
$t_G^{\alpha_i}$, $i=1,\dots,M$ are the
$SU(3)_{color}$ group generators (in the adjoint representation) with $Tr_C$
the trace over the color group, the $u_i$ are Feynman parameters, the
function $\theta$ is specified by
\newline \mbox{$\theta(u_M,\ldots,u_1)=\theta(u_M-u_{M-1})\ldots\theta(u_2-u_1)$}
and
\[
F^{(M)}(u_1,\ldots,u_M;T)=
\left[\prod_{n=M}^1 \int d\xi_n d\bar{\xi}_n\right]
\left(Tr_L \Phi^{[1]} -2 \right) \times \hspace{5cm}
\]
\begin{equation}
\times \exp\left[\sum_{n \neq m} \xi_n \bar{\xi}_n
\varepsilon^n \cdot p_m \partial_n G(u_n,u_m)+
\frac{1}{2T} \sum_{n \neq m} \xi_n \bar{\xi}_n \xi_m \bar{\xi}_m
\varepsilon^n \cdot \varepsilon^m \partial_n \partial_m G(u_n,u_m)\right]\;,
\end{equation}

In the above equation $\xi$'s are Grassmann variables,  the $\varepsilon^i$ are 
polarization vectors for the external gluons, $\Phi^{[1]}$ is the so-called {\it spin factor} 
entering the world-line description of QCD (see below), with $Tr_L$ denoting trace with 
respect to Lorenz generator representation indices and the $G(u_n,u_m)$ are free 
propagators for the particle modes entering the worldline path integral description of 
QCD, in the context of its first quantized version (see Ref [9]), obeying the equation(s)
\begin{equation}
-\partial_n\partial_m G(u_n,u_m) = \partial_n^2\dot{G}(u_n,u_m) \equiv
\ddot{G}(u_n,u_m)=2[\delta(u_n,u_m)-1]\;,
\end{equation}
with boundary condition
\begin{equation}
\partial_n G(u_n,u_m) \equiv \dot{G}(u_n,u_m)=
sign(u_n-u_m)-2(u_n-u_m)=-\dot{G}(u_m,u_n)\;,
\end{equation}
The explicit expression for the spin factor in terms of the set of parameters
entering our expressions is (the $J_{\mu\nu}$ are the Lorentz generators, in
the vector representation)
\[
\Phi_{\mu\nu}^{[1]}=
P\exp\left[\frac{i}{2}\sum_{n=1}^{M} J \cdot \phi(n) \right]_{\mu\nu} =
\hspace{10cm}
\]
\begin{equation}
=\delta_{\mu\nu}+\frac{i}{2}(J_{\rho\sigma})_{\mu\omega}
\sum_{n=1}^{M}\phi_{\rho\sigma}(n)+
(\frac{i}{2})^2 (J_{\rho_2\sigma_2})_{\mu\lambda}
(J_{\rho_1\sigma_1})_{\lambda\nu}
\sum_{n_2=1}^M \sum_{n_1=1}^{n_2-1}
\phi_{\rho_2\sigma_2}(n_2) \phi_{\rho_1\sigma_1}(n_1) + \ldots \;,
\end{equation}
where
\begin{equation}
\phi_{\mu\nu}(n)=2\bar{\xi}_n\xi_n
(\varepsilon_\mu^n p_{n,\nu}-\varepsilon_\nu^n p_{n,\mu})
+\frac{4}{T} \bar{\xi}_{n+1}\xi_{n+1} \bar{\xi}_n\xi_n
(\varepsilon_\mu^{n+1} \varepsilon_\nu^n -
 \varepsilon_\nu^{n+1} \varepsilon_\mu^n ) \delta(u_{n+1}-u_n)\;.
\end{equation}

A point of note is the following: In the above expressions a specific time
ordering has been chosen according to which index $n+1$ comes immediately
after index $n$, with $\xi_{M+1}=\bar{\xi}_{M+1}=0$.

\vspace{3cm}
{\large \bf 3. Presenting the program structure}

The form of eq. (1) is considered as a sum of different elements. Every element
is described with a line in three matrices: $C1$, $C2$ and $C3$.

Matrix $C1$ contains the Grassmann part of the form. It is a matrix with
integer elements with dimensions $(NC1 \times NC2)$\footnote{The first number
is the number of lines and the second the number of columns.}.

Matrix $C2$ has dimensions $(NC1 \times 1)$ and contains real numbers.

Matrix $C3$ represents the functions that accompany the Grassmann variables
and its dimensions are $(NC1 \times NC3)$. The representation of the functions
will be discussed in the ``Functions Handling'' subsection.

\vspace{0.2cm}
{\it 3a. Grassmann Coding}

A Grassmann variable is represented in matrix $C1$ by an integer number. This
integer is equal to the index of the Grassmann number for the $\xi$ type
variables and equal to the opposite of the index of the Grassmann number for
the $\bar{\xi}$ type variables. So
\begin{equation}
\xi_n\;\;\rightarrow\;\;n\;,\;\;\;\;\bar{\xi}_n\;\;\rightarrow\;\;-n\;.
\end{equation}

A form that contains products of Grassmann variables is represented by a line
in matrix $C1$. The elements in this line are put in the same order by which
the Grassmann variables enter a given product. This line is accompanied by a
line in matrix $C2$ which
contains a real number and a line matrix $C3$ which represents the functions.
For example the following coding would take place
\begin{equation}
\alpha \cdot func_1 \cdot func_2 \cdot \xi_1\bar{\xi}_1\xi_5\bar{\xi}_5
\;\rightarrow\;
\stackrel{C2}{(\begin{array}{c}\alpha\end{array})}
\stackrel{C3}{(\begin{array}{cc}f_1&f_2\end{array})}
\stackrel{C1}{(\begin{array}{cccc}1&-1&5&-5\end{array})}\;,
\end{equation}
where the integers $f_1,f_2$ represent the functions $func_1,func_2$ and will
be discussed in the following subsection.

The addition of two forms containing Grassmann variables (represented as
$\oplus$) is carried as follows
\[
\alpha_1 \cdot func_1 \cdot func_2 \cdot \xi_2\bar{\xi}_2 +
\alpha_2 \cdot func_3 \cdot \xi_3\bar{\xi}_3 \xi_1\bar{\xi}_1
\rightarrow
\]
\[
\stackrel{C2}{(\begin{array}{c}\alpha_1\end{array})}
\stackrel{C3}{(\begin{array}{cc}f_1&f_2\end{array})}
\stackrel{C1}{(\begin{array}{cc}2&-2\end{array})}\oplus
\stackrel{C2}{(\begin{array}{c}\alpha_2\end{array})}
\stackrel{C3}{(\begin{array}{c}f_3\end{array})}
\stackrel{C1}{(\begin{array}{cccc}3&-3&1&-1\end{array})}
\;\rightarrow
\]
\begin{equation}
\stackrel{C2}{\left( \begin{array}{r}\alpha_1\\ \alpha_2 \end{array} \right)}
\stackrel{C3}{\left( \begin{array}{rr}f_1&f_2\\ f_3&0 \end{array} \right)}
\stackrel{C1}{\left( \begin{array}{rrrr}2&-2&0&0\\ 3&-3&1&-1\end{array}
\right)}\;.
\end{equation}
The result of the addition of two forms represented with two sets of matrices
$C2$, $C3$ and $C1$ will be a new set of these matrices with new dimensions. If
the dimensions of the first set of matrices are represented by $N$'s and the
these of the second set by $M$'s, then the resulting dimensions of the final
set will be
\[
\stackrel{C2}{\left(N_1 \times 1\right)}
\stackrel{C3}{\left(N_1 \times N_3\right)}
\stackrel{C1}{\left(N_1 \times N_2\right)} \oplus
\stackrel{C2}{\left(M_1 \times 1\right)}
\stackrel{C3}{\left(M_1 \times M_3\right)}
\stackrel{C1}{\left(M_1 \times M_2\right)} \;\rightarrow
\]
\begin{equation}
\stackrel{C2}{\left[(N_1+M_1) \times 1\right]}
\stackrel{C3}{\left[(N_1+M_1) \times \max(N_3,M_3)\right]}
\stackrel{C1}{\left[(N_1+M_1) \times \max(N_2,M_2)\right]} \;.
\end{equation}
So the dimensions of the matrices which represent our forms is not constant
and it is important for every part of the program to be known.
Where it is necessary, the lines of matrices $C3$ and $C1$ are filled with
zeros. The addition of two forms is carried out in the program by the
subroutine ``ADD''.

The multiplication of two forms containing Grassmann variables (represented as
$\otimes$) is carried as follows
\[
\left(\alpha_1 \cdot func_1 \cdot func_2 \cdot \xi_2\bar{\xi}_2 +
\alpha_2 \cdot func_3 \cdot \xi_3\bar{\xi}_3 \xi_1\bar{\xi}_1\right)\cdot
\left(\beta_1 \cdot func_4 \cdot \xi_5\bar{\xi}_5 +
\beta_2 \cdot func_5 \cdot \xi_7\bar{\xi}_7 \right)
\rightarrow
\]
\[
\stackrel{C2}{\left( \begin{array}{r}\alpha_1\\ \alpha_2 \end{array} \right)}
\stackrel{C3}{\left( \begin{array}{rr}f_1&f_2\\ f_3&0 \end{array} \right)}
\stackrel{C1}{\left( \begin{array}{rrrr}2&-2&0&0\\ 3&-3&1&-1\end{array} \right)}
\otimes
\stackrel{C2}{\left( \begin{array}{r}\beta_1\\ \beta_2 \end{array} \right)}
\stackrel{C3}{\left( \begin{array}{r}f_4\\ f_5 \end{array} \right)}
\stackrel{C1}{\left( \begin{array}{rr}5&-5\\ 7&-7\end{array}\right)}
\;\rightarrow
\]
\begin{equation}
\stackrel{C2}{\left( \begin{array}{r}
\alpha_1\cdot\beta_1 \\ \alpha_1\cdot\beta_2\\
\alpha_2\cdot\beta_1 \\ \alpha_2\cdot\beta_2 \end{array} \right)}
\stackrel{C3}{\left( \begin{array}{rrr}
f_1&f_2&f_4 \\ f_1&f_2&f_5 \\ f_3&0&f_4  \\ f_3&0&f_5 \end{array} \right)}
\stackrel{C1}{\left( \begin{array}{rrrrrr}
2&-2&0&0&5&-5 \\ 2&-2&0&0&7&-7 \\ 3&-3&1&-1&5&-5 \\ 3&-3&1&-1&7&-7
\end{array} \right)}\;.
\end{equation}

The result of the multiplication of two forms represented with two sets of matrices
$C2$, $C3$ and $C1$ will be a new set of these matrices with new dimensions. If
the dimensions of the first set of matrices are represented by $N$'s and
those of the second set by $M$'s, then the resulting dimensions of the final
set will be
\[
\stackrel{C2}{\left(N_1 \times 1\right)}
\stackrel{C3}{\left(N_1 \times N_3\right)}
\stackrel{C1}{\left(N_1 \times N_2\right)} \otimes
\stackrel{C2}{\left(M_1 \times 1\right)}
\stackrel{C3}{\left(M_1 \times M_3\right)}
\stackrel{C1}{\left(M_1 \times M_2\right)} \;\rightarrow
\]
\begin{equation}
\stackrel{C2}{\left[(N_1\cdot M_1) \times 1\right]}
\stackrel{C3}{\left[(N_1\cdot M_1) \times (N_3+M_3)\right]}
\stackrel{C1}{\left[(N_1\cdot M_1) \times (N_2+M_2)\right]} \;.
\end{equation}
The multiplication of two forms is carried out in the program by the
subroutine ``MULTIPLY''.

Certain simplifications and rearrangements are carried out in the Grassmann
and Function Number matrices upon the completion of certain routines. This is
done by subroutine ``REARRANGE'' which performs the following acts:

$\bullet$ It checks if in the same line two equal Grassmann variables exist and
if that happens it puts zeros everywhere in the lines of matrices $C2$, $C3$
and $C1$. This accounts for the fact that $\xi_n^2$=$\bar{\xi}_n^2$=0.

$\bullet$ If a zero exists in one line of matrix $C2$, then the corresponding
lines in all matrices are omitted, all the following lines are moved in front
by one line and the line dimension is reduced by one.

$\bullet$ It collects the Grassmann variables to the left of every line of matrix $C1$,
ignoring the zeros in between. If this reduces the dimension of the columns
of matrix $C1$, it performs the reduction. The same act is performed on
function matrix $C3$.

$\bullet$ It places the integers which represent the Grassmann variables in
every line of matrix $C1$ in ascending order, according to their absolute
value. Among two variables with the same absolute value the positive is placed
to the left of the negative one. The sign of the corresponding value of matrix
$C2$ is changed according to the changes made to the order of the Grassmann
variables.

$\bullet$ It places the number which represents the functions in every line of
matrix $C3$ in ascending order. The set of functions which are not allowed to
change their relative order are placed to the left. Functions represented by
numbers $3nm$ with $n>m$ are entered as $3mn$. This practice, as well as the two
preceding ones, are needed for comparing the different lines of the matrices.

$\bullet$ If two lines in the matrix $C1$ and the corresponding lines in the
matrix $C3$ are found equal, then the accompanying factors of matrix $C2$ are added.
The second line is omitted in all matrices and their line dimension is reduced
by one.

$\bullet$ The lines of the matrices are also rearranged. The lines with all
elements of matrices $C1$ and $C3$ equal to zero are entered first.

The matrices $C3$ and $C1$ are two-dimensional, but their dimension is not
known {\it a priori}. This can cause problems when the matrices are fed from
one subroutine to another. In every subroutine the dimensions of these
matrices are pre-defined so that they will always be larger than what is required.
Also, it is not always possible to define the same dimensions for these
matrices in all the subroutines. The FORTRAN language stores the elements of
a two-dimensional matrix in neighbouring memory places so that the first column is
stored first, then the second, etc. When these places are read from another
subroutine they will not result to the same matrix in the new subroutine,
unless the matrix has the same dimensions. To avoid this problem 
the subroutine ``LINEUP'' is used, which rearranges the elements of a matrix
that is defined with dimensions $(N_1\times N_2)$ so that they are stored
exactly the same way as they would be stored if the dimensions of the matrix
were $(N_1'\times N_2')$. In other cases we deal with this problem by
using one-dimensional arrays to store two-dimensional matrices. The elements
of the matrix $a_{ij}$, with dimensions $(N_1\times N_2)$, are stored into the
elements $b$ of the array, so that the first column of the matrix is stored
into the first elements of the array, then the second, etc., according to the
correspondence
\begin{equation}
a_{ij}\;\;\rightarrow\;\;b_{(j-1)\cdot N_1+i}\;\;.
\end{equation}

To evaluate eq. (1) we first code the following form in matrices $C2$, $C3$
and $C1$
\[
T\sum_{n<m} p_n \cdot p_m G(u_n,u_m) +
\sum_{n \neq m} \xi_n \bar{\xi}_n
\varepsilon^n \cdot p_m \partial_n G(u_n,u_m)+
\]
\begin{equation}
\frac{1}{2T} \sum_{n \neq m} \xi_n \bar{\xi}_n \xi_m \bar{\xi}_m
\varepsilon^n \cdot \varepsilon^m \partial_n \partial_m G(u_n,u_m)
\delta(u_{n+1}-u_n)\;.
\end{equation}
This is done by subroutine ``KNKM''. Then the exponent of the above form has
to be evaluated. For this purpose the subroutine ``EXPONENTIAL'' has been
constructed. In this routine it is checked whether a part that does not
contain Grassmann variables exists. If this is the case, the exponent of this
part will only multiply all the lines of the rest. Then the exponent of the
Grassmann containing part is calculated. This exponent will only contain $M$ terms,
since there are only $M$ different Grassmann pairs $\xi_n\bar{\xi}_n$. Every
term is constructed by multiplying the Grassmann part of ``KNKM'' as many times
as it is needed. Every newly constructed term is then added to the pervious
ones. The result is stored to the output of subroutine ``EXPKNKM''.

The structure $(i/2) J \cdot \phi(n)$ is coded in subroutine
``FMN''. The index $n+1$ is evaluated from the specific time order which has
been chosen. The exponent of ``FMN'' is found by subroutine ``PEXPONENTIAL''.
The trace of $\delta_{\mu\nu}$ (the first term in the sum of eq. (6)) is
accounted for by putting the factor $4$ to a line of matrix $C2$.

The complete result is determined in subroutine ``MULTIPLYALL''. In the
output of ``PEXPONENTIAL'' a line with zeros in the domain of matrices $C3$
and $C1$ but with the factor $-2$ in $C2$ is added, to construct the form
$Tr_L \Phi^{[1]} -2$. Then the result is
multiplied with the result of ``EXPKNKM''. Now the complete Grassmann part of
the calculation is stored to matrix $C1$. The Grassmann integration is simple,
since it amounts to keeping only the terms where all the Grassmann variables
are present. Another simplification is made by neglecting the terms
that contain only one function to be traced\footnote{Function of the type $2n$
or $3n$ of the next subsection.} and which has zero trace. Our calculation at
the end of ``MULTIPLYALL'' is stored only in matrices $C2$ and $C3$ since the
Grassmann variables have been integrated out.

\vspace{0.2cm}
{\it 3b. Function Handling}

In order to keep the matrices that represent our forms as small as possible,
we have first carried out the Grassmann integrations in order to remove
completely the Grassmann variables. In fact, this is the reason we have
coded the functions that accompany the Grassmann variables in matrix $C3$.
Every line of this matrix contains a product of functions which are
represented by integer numbers according to the labelling described in
Table 1.

In order for the representation of Table 1 to be valid it must be $M\leq9$.
The asterisk has the meaning that the corresponding functions are considered
non-commutative and the order that they have been inserted is not allowed to
change.
Their Lorentz indices $\mu$, $\nu$ are also considered indefinite, that is
they are not defined at this point by the number which represents the
function. So when in a line of matrix $C3$ functions of the type $2n$ or $3n$
are found together it is assumed that the Lorentz trace has to be performed,
according to the exact order the functions are found in the respective line.
For example if the numbers $21$, $22$ and $31$ are found in one line of $C3$
(placed in this order and without any other function of the type $2n$ or $3n$ in
the same line), then it is assumed that
\[
21\;22\;31\;\rightarrow\;
(21)_{\mu\rho}\;(22)_{\rho\sigma}\;(31)_{\sigma\mu}\;\rightarrow
\]
\[
(i/2)(J_{\alpha\beta})_{\mu\rho} 2
(\varepsilon_\alpha^1 p_{1,\beta}-\varepsilon_\beta^1 p_{1,\alpha})
(i/2)(J_{\gamma\delta})_{\rho\sigma} 2
(\varepsilon_\gamma^2 p_{2,\delta}-\varepsilon_\delta^2 p_{2,\gamma})
\]
\begin{equation}
(i/2)(J_{\kappa\lambda})_{\sigma\mu} (1/T)
(\varepsilon_\kappa^2 \varepsilon_\lambda^1-
 \varepsilon_\lambda^2 \varepsilon_\kappa^1)
\end{equation}
In this way we postpone the evaluation of the trace until after the Grassmann
integration.

After the Grassmann integration we are able to perform the trace in the
surviving terms. This is accomplished by subroutine ``TRACE''.
The information contained in matrix $C3$ is divided into two matrices $C3$ and
$C6$. The new matrix $C3$ contains only the following functions according to
the labelling

Matrix $C6$ contains in its first column the exponent of $T$ followed by
internal products of the vectors $\varepsilon$'s and $p$'s, according to the
coding of Table 1.

In order to arrive at that output, the maximum number of traces to be
calculated is evaluated first, so that the
dimension of the matrix that will contain the result is determined.
The saturation of indices $\rho$, $\sigma$ in $J_{\rho\sigma}$ is performed
instantly, since
\begin{equation}
(i/2)(J_{\rho\sigma})_{\mu\nu} 2
(\varepsilon_\rho^n p_{n,\sigma}-\varepsilon_\sigma^n p_{n,\rho})=
-2(\varepsilon_\mu^n p_{n,\nu}-\varepsilon_\nu^n p_{n,\mu})\;,
\end{equation}
\begin{equation}
(i/2)(J_{\rho\sigma})_{\mu\nu} (1/T)
(\varepsilon_\rho^{n+1} \varepsilon_\sigma^n-
 \varepsilon_\sigma^{n+1} \varepsilon_\rho^n)=
-(1/T)(\varepsilon_\mu^{n+1} \varepsilon_\nu^n-
 \varepsilon_\nu^{n+1} \varepsilon_\mu^n)
\end{equation}

Then the indefinite Lorentz indices become definite. We begin by giving the
indices 1,2 to the first function that enters the product to be traced, then
the indices 2,3 to the second, etc, until the last one which takes
as second index 1. The representation used is
\begin{equation}
\varepsilon_\nu^n\;\rightarrow \;3n\nu\;,\;\;\; p_{n,\nu}\rightarrow \;2n\nu\;.
\end{equation}
The products are then represented as lines in two temporary matrices and
every term in a sum uses a separate line, for example
\begin{equation}
\varepsilon_1^1 p_{2,2}-\varepsilon_2^1 p_{2,1}\;\rightarrow
\left(\begin{array}{rr}1\\-1\end{array}\right)
\left(\begin{array}{rr}311&222\\312&221\end{array}\right)\;.
\end{equation}
The addition and the multiplication of forms as the above is carried out in the
same way as was done when we dealt with the Grassmann variables, until all
the terms to enter the product to be traced are exhausted. In order to form
internal products of the different elements, the last digit of every element
is compared to the last digit of all the elements to the right in the same
line. When two last digits are found equal an internal product between the
corresponding elements is formed and represented according to the prescription
of Table 3.

We have, also, taken into account the rule $\varepsilon^n \cdot p_n = 0$ and
we have made everywhere in matrix $C6$ (but not in $C3$) the replacement
$p_1=-p_2-p_3 \ldots -p_M$, according to momentum conservation.

The result that corresponds to one line of matrix $C3$ is put in one line of
matrix $C6$. This contains a sum of different terms, each of which is a
product of internal products accompanied by a factor. When the line of matrix
$C6$ is read, an integer less or equal to 2000 (which represents a factor)
signals the beginning of the product of internal products and a another one
signals its ending. For example the following translation is assumed from
a line of matrix $C6$
\[
\begin{array}{cccccccccc}
-1&3431&1&3433&3223&3122&2424&-3&3423&3332 \end{array} \; \rightarrow
\]
\begin{equation}
-\varepsilon^4\cdot\varepsilon^1+
\varepsilon^4\cdot \varepsilon^3\;\varepsilon^2\cdot p_3
\;\varepsilon^1\cdot p_2\;p_4\cdot p_4
-3\;\varepsilon^4\cdot p_3\;\varepsilon^3\cdot \varepsilon^2\;.
\end{equation}

\vspace{0.3cm}
{\large \bf 4. Results}

At the beginning of its run the program asks the user to insert the
number of external gluons $M$. Then it calculates all the possible
combinations of time order and lists them for the user to see. Every
combination is accompanied by an integer number and the user is expected to
insert one of these numbers in order to  choose the desirable combination.

Then the calculation is carried out. Several files are used during this
calculation. The results after the Grassmann integration (matrices $C2$ and
$C3$ in coding of Table 1) are written in file ``TEMP2''. The final
result (matrices $C2$, $C3$ in coding of Table 2 and $C6$ in coding of
integers) are written in file ``TEMP4''. 
The result in function like output is written to file ``TEMP5''.
The integers NC1, NC3 and NC6 are connected
to the dimensions of the output matrices. These are $C2(NC1\times 1)$,
$C3(NC1\times NC3)$ and $C6(NC1\times NC6)$.
The result is read as a sum of different terms. Every term is formed by
reading a line of matrix $C2$, then the corresponding line of matrix $C3$ and
then the corresponding line of matrix $C6$. When a lot of non-zero elements
exist in some lines of matrix $C6$, these lines are folded beneath in order
to permit easy reading. Then the whole sum is supposed to be multiplied by
the function and integrated by the multiple integral which appear at the
begging of the output.

The number of terms at the output ($NC1$) grows rapidly when $M$ increases.
For $M=2$, three terms exist (shown to Table 4a). For $M=3$, 23 terms exist.
These terms are shown in Table 4b. For $M=4$ the existing terms are 233.
These include both finite and divergent terms.

Our results have been compared with the analytic calculations for the
divergent terms for $M=2$, $M=3$ and $M=4$ [9] and have been found in total
agreement.
The structure of the program and the methods used for the computation of the
master formulas permit the extension of the calculations for value of $M$
higher than 4.

\vspace{0.3cm}
{\large \bf 5. Concluding remarks}

In this paper a computational algorithm has been presented for the successful
computation of the complete set of one-gluon loop Feynman diagrams with two,
three and four external gluon attachments, on the basis of the master formulas
derived in Ref [9]. It is hoped that the particular feature of the constructed
algorithm, namely the ability to expedite integrations over a multivariable
set, a subset of which is Grassmannian, could find wider applications to
analogous situations that may arise in other physical problems wherein
Grassmann variables make their entrance. Within the context of the present
application, it would be of interest to apply the particular algorithm
developed in this work to the two-gluon loop $M=4$ case, the corresponding
master expressions for which have been derived in Ref [10]. As a first
attempt, one could restrict the relevant computation to the divergent term
associated with the $M=2$ configuration and verify the consistency with
second order corrections to the running coupling constant in pQCD.

\vspace{0.3cm}
{\large \bf Acknowledgement}

Two of the present authors, A. I.  K. and C. N. K. acknowledge the support
from the General Secretariat of Research and Technology of the
University of Athens.

\newpage
{\large \bf Tables}
\vspace{0.3cm}
\begin{center}
\begin{tabular}{cl} \hline
Function & Number representing \\
         & Function \\ \hline
$\displaystyle (i/2)(J_{\rho\sigma})_{\mu\nu} 2
(\varepsilon_\rho^n p_{n,\sigma}-\varepsilon_\sigma^n p_{n,\rho})$ & 2n (*) \\
$\displaystyle (i/2)(J_{\rho\sigma})_{\mu\nu} (1/T)
(\varepsilon_\rho^{n+1} \varepsilon_\sigma^n-
 \varepsilon_\sigma^{n+1} \varepsilon_\rho^n) \delta(u_{n+1}-u_n)$ & 3n (*) \\
$\displaystyle \exp \{T \sum_{n=1}^{M} \sum_{m=n+1}^{M} [p_n \cdot p_m G(u_n,u_m)]\}$ & 100 \\
$\displaystyle \varepsilon^n \cdot p_m \; \partial_n G(u_n,u_m)$ & 2nm \\
$\displaystyle (1/2T) \varepsilon^n \cdot \varepsilon^m \partial_n \partial_m G(u_n,u_m)$ & 3nm=3mn \\ \hline
\end{tabular} 
\end{center} 

\begin{center}
{\it Table 1. The representation of functions through integers before the
Grassmann integration in matrix $C3$.}
\end {center}

\vspace{0.3cm}
\begin{center}
\begin{tabular}{cl} \hline
Function & Number representing \\
         & Function \\ \hline
$\displaystyle \delta(u^{n+1}-u^n)$ & 3n \\
$\displaystyle \exp \{T \sum_{n=1}^{M} \sum_{m=n+1}^{M}
[p_n \cdot p_m G(u_n,u_m)]\}$ & 100 \\
$\displaystyle \varepsilon^n \cdot p_m \; \partial_n G(u_n,u_m)$ & 2nm \\
$\displaystyle \varepsilon^n \cdot \varepsilon^m \partial_n \partial_m G(u_n,u_m)$ & 3nm=3mn \\ \hline
\end{tabular} 
\end{center} 

\begin{center}
{\it Table 2. The representation of functions through integers in matrix $C3$
at the output.}
\end {center}

\vspace{0.3cm}
\begin{center}
\begin{tabular}{cc} \hline
Function & Number representing \\
         & Function \\ \hline
$p_n \cdot p_m$ & 2n2m \\
$\varepsilon^n \cdot p_m$ & 3n2m \\
$\varepsilon^n \cdot \varepsilon^m$ & 3n3m \\ \hline
\end{tabular} 
\end{center} 

\begin{center}
{\it Table 3. The representation of functions through integers in matrix $C6$
at the output.}
\end {center}

\renewcommand{\baselinestretch}{1.2}
{\scriptsize

\underline{\hspace{12.3cm}}

M=           2

NC1,NC3,NC6=           3           3           4

TIME ORDER=      1      2

\vspace{0.2cm}
-(pi**2/2)*g**(2)*dl4(p1+p2 )*TrC(tGa2 tGa1 )*

\vspace{0.2cm}
\begin{tabular}{ccccccccc}
 inf    &    & 1      &     & u2     &     &            \\
 $\mid$ & dT & $\mid$ & du2 & $\mid$ & du1 &  th(u2,u1 )\\
 0      &    & 0      &     & 0      &     &            \\
\end{tabular}

\vspace{0.5cm}
\begin{tabular}{rrrrrrrrrrrr} 
 4.0&  EXP(all)&                  &                  & T**(-1)&   +2& e2.e1&p2.p2 \\ 
 2.0&  EXP(all)& e1.p2d1G(u1,u2)  & e2.p1d2G(u2,u1)  & T**(-1)&     &      &      \\
 2.0&  EXP(all)& e1.e2d1d2G(u1,u2)&                  & T**(-2)&     &      &      \\ \hline
\end{tabular}

}
\renewcommand{\baselinestretch}{1.5}

\begin{center}
{\it Table 4a. The function like output of the program for $M=2$ and for the
time order $u_1<u_2$. The Memoradum is similar to Table 4b.}
\end {center}

\newpage
\renewcommand{\baselinestretch}{1.2}
{\scriptsize

\hspace{-1.5cm}\underline{\hspace{18.6cm}}

\hspace{-1.5cm} M=          3

\hspace{-1.5cm} NC1,NC3,NC6=         23           4          41

\hspace{-1.5cm} TIME ORDER=          1           2           3

\vspace{0.2cm}
\hspace{-1.5cm} -(pi**2/2)*g**(3)*dl4(p1+p2+p3 )*TrC(tGa3 tGa2 tGa1 )*

\vspace{0.2cm}
\hspace{-1.5cm}
\begin{tabular}{ccccccccc}
 inf    &    & 1      &     & u3     &     & u2     &     &               \\
 $\mid$ & dT & $\mid$ & du3 & $\mid$ & du2 & $\mid$ & du1 &  th(u3,u2,u1 )\\
 0      &    & 0      &     & 0      &     & 0      &     &               \\
\end{tabular}

\vspace{0.5cm}
\hspace{-1.5cm}
\begin{tabular}{rrrrrrrrrrrr} 
-8.0&  EXP(all)&                &                  &                & T**( 0)&   -1& e3.p2&e2.p3&e1.p2&     &      \\ 
    &          &                &                  &                &        &   +1& e3.e2&e1.p2&p3.p2&     &      \\
    &          &                &                  &                &        &   +1& e3.e2&e1.p2&p3.p3&     &      \\
    &          &                &                  &                &        &   -1& e3.p2&e2.e1&p3.p3&     &      \\
    &          &                &                  &                &        &   +1& e3.e1&e2.p3&p2.p2&     &      \\
    &          &                &                  &                &        &   +1& e3.e1&e2.p3&p2.p3&     &      \\
    &          &                &                  &                &        &   -1& e3.e2&e1.p3&p2.p2&     &      \\
    &          &                &                  &                &        &   -1& e3.e2&e1.p3&p2.p3&     &      \\
    &          &                &                  &                &        &   -1& e3.e1&e2.p3&p3.p2&     &      \\
    &          &                &                  &                &        &   +1& e3.p2&e2.p3&e1.p3&     &      \\
 4.0&  EXP(all)& e1.p2d1G(u1,u2)&                  &                & T**( 0)&   +2& e3.p2&e2.p3&   -2&e3.e2&p3.p2 \\
 4.0&  EXP(all)& e1.p3d1G(u1,u3)&                  &                & T**( 0)&   +2& e3.p2&e2.p3&   -2&e3.e2&p3.p2 \\
 4.0&  EXP(all)& e2.p1d2G(u2,u1)&                  &                & T**( 0)&   -2& e3.p2&e1.p3&   +2&e3.e1&p3.p2 \\
    &          &                &                  &                &        &   +2& e3.e1&p3.p3&     &     &      \\  
 4.0&  EXP(all)& e2.p3d2G(u2,u3)&                  &                & T**( 0)&   -2& e3.p2&e1.p3&   +2&e3.e1&p3.p2 \\
    &          &                &                  &                &        &   +2& e3.e1&p3.p3&     &     &      \\  
 4.0&  EXP(all)& e3.p1d3G(u3,u1)&                  &                & T**( 0)&   -2& e2.p3&e1.p2&   +2&e2.e1&p2.p2 \\
    &          &                &                  &                &        &   +2& e2.e1&p2.p3&     &     &      \\  
 4.0&  EXP(all)& e3.p2d3G(u3,u2)&                  &                & T**( 0)&   -2& e2.p3&e1.p2&   +2&e2.e1&p2.p2 \\
    &          &                &                  &                &        &   +2& e2.e1&p2.p3&     &     &      \\  
 2.0&  EXP(all)& e1.p2d1G(u1,u2)& e2.p1d2G(u2,u1)  & e3.p1d3G(u3,u1)& T**( 0)&     &      &     &     &     &      \\
 2.0&  EXP(all)& e1.p2d1G(u1,u2)& e2.p1d2G(u2,u1)  & e3.p2d3G(u3,u2)& T**( 0)&     &      &     &     &     &      \\   
 2.0&  EXP(all)& e1.p2d1G(u1,u2)& e2.p3d2G(u2,u3)  & e3.p1d3G(u3,u1)& T**( 0)&     &      &     &     &     &      \\   
 2.0&  EXP(all)& e1.p2d1G(u1,u2)& e2.p3d2G(u2,u3)  & e3.p2d3G(u3,u2)& T**( 0)&     &      &     &     &     &      \\   
 2.0&  EXP(all)& e1.p3d1G(u1,u3)& e2.p1d2G(u2,u1)  & e3.p1d3G(u3,u1)& T**( 0)&     &      &     &     &     &      \\   
 2.0&  EXP(all)& e1.p3d1G(u1,u3)& e2.p1d2G(u2,u1)  & e3.p2d3G(u3,u2)& T**( 0)&     &      &     &     &     &      \\   
 2.0&  EXP(all)& e1.p3d1G(u1,u3)& e2.p3d2G(u2,u3)  & e3.p1d3G(u3,u1)& T**( 0)&     &      &     &     &     &      \\   
 2.0&  EXP(all)& e1.p3d1G(u1,u3)& e2.p3d2G(u2,u3)  & e3.p2d3G(u3,u2)& T**( 0)&     &      &     &     &     &      \\   
 8.0&  EXP(all)& dl(u3-u2)      &                  &                & T**(-1)&   -2& e3.p2&e2.e1&   +2&e3.e1&e2.p3 \\
 8.0&  EXP(all)& dl(u2-u1)      &                  &                & T**(-1)&   +2& e3.e1&e2.p3&   -2&e3.e2&e1.p3 \\
 2.0&  EXP(all)& e1.p2d1G(u1,u2)& e2.e3d2d3G(u2,u3)&                & T**(-1)&     &      &     &     &     &      \\  
 2.0&  EXP(all)& e1.p3d1G(u1,u3)& e2.e3d2d3G(u2,u3)&                & T**(-1)&     &      &     &     &     &      \\   
 2.0&  EXP(all)& e2.p1d2G(u2,u1)& e1.e3d1d3G(u1,u3)&                & T**(-1)&     &      &     &     &     &      \\   
 2.0&  EXP(all)& e2.p3d2G(u2,u3)& e1.e3d1d3G(u1,u3)&                & T**(-1)&     &      &     &     &     &      \\   
 2.0&  EXP(all)& e3.p1d3G(u3,u1)& e1.e2d1d2G(u1,u2)&                & T**(-1)&     &      &     &     &     &      \\  
 2.0&  EXP(all)& e3.p2d3G(u3,u2)& e1.e2d1d2G(u1,u2)&                & T**(-1)&     &      &     &     &     &      \\ \hline   

\end{tabular}

\hspace{-1.5cm}Memorandum:

\hspace{-1.5cm}\mbox{
\begin{tabular}{ll|ll|ll}
pi**2          & $\pi^2$                   &
th(u3,u2,u1 )  & $\theta (u_3,u_2,u_1)$    &
e1.p2          & $\varepsilon^1 \cdot p_2$
\\
dl4(p1+p2+p3 ) & $\delta^{(4)}(p_1+p_2+p_3)$                                                           & 
EXP(all)       & $\displaystyle \exp \{T \sum_{n=1}^{M} \sum_{m=n+1}^{M} [p_n \cdot p_m G(u_n,u_m)]\}$ &
e2.e3          & $\varepsilon^2 \cdot \varepsilon^3$
\\
TrC(tGa3 tGa2 tGa1 ) & $Tr_C (t_G^{\alpha_3} t_G^{\alpha_2} t_G^{\alpha_1})$ &
d2d3G(u2,u3)         & $\partial_2 \partial_3 G(u_2,u_3)$                    &
d1G(u1,u2)           & $\partial_1 G(u_1,u_2)$
\\
inf         & $\infty$            & 
dl(u3-u2)   & $\delta(u_3 - u_2)$ &
T**(-1)     & $T^{-1}$
\\
\end{tabular}
}

}
\renewcommand{\baselinestretch}{1.5}

\begin{center}
{\it Table 4b. The function like output of the program for $M=3$ and for the
time order $u_1<u_2<u_3$.}
\end {center}

{\large{\bf Figure Caption}}\\
\rm {\bf Figure} Illustration, for $M=4$, of the classes of one-gluon loop
Feynman diagram (right side of arrow) which are simultaneously accommodated by
the corresponding master formula depicted on left side of arrow.


\begin{thebibliography}{99}

\bibitem{Bern:1988pk} Z.~Bern and D.~A.~Kosower,
Nucl.\ Phys.\ B {\bf 321} (1989)605;  B {\bf 379} (1992)451.

\bibitem{Bern:1993mq} Z.~Bern, L.~J.~Dixon and D.~A.~Kosower,
Phys.\ Rev.\ Lett.\ {\bf 70} (1993)2677 [arXiv:hep-ph/9302280].

\bibitem{DiVecchia:1996uq} P.~Di Vecchia, L.~Magnea, A.~Lerda, R.~Russo
and R.~Marotta,
Nucl.\ Phys.\ B {\bf 469} (1996)235 [arXiv:hep-th/9601143].

\bibitem{Strassler:1992zr} M.~J.~Strassler,
Nucl.\ Phys.\ B {\bf 385} (1992)145 [arXiv:hep-ph/9205205].

\bibitem{Reuter:1997zm} M.~Reuter, M.~G.~Schmidt and C.~Schubert,
Annals Phys.\ {\bf 259} (1997)313 [arXiv:hep-th/9610191].

\bibitem{Sato:1998kq}  H-T. Sato and M. G. Schmidt, Nucl.\ Phys.\ B {\bf
524} (1998)742;
{it ibid.} {\bf 560} (1999)551.

\bibitem{Sato:2000cr}  H-T. Sato, M. G. Schmidt and Claus Zahlten,
Nucl.\ Phys.\ B {\bf 579} (2000)492 [arXiv: hep-th/0003070].

\bibitem{Karanikas:1999hz} A.~I.~Karanikas and C.~N.~Ktorides,
JHEP {\bf 9911} (1999)033 [arXiv:hep-th/9905027].

\bibitem{Avramis:2002xf} S.~D.~Avramis, A.~I.~Karanikas and
C.~N.~Ktorides,
Phys.\ Rev.\ D {\bf 66} (2002)045017 [arXiv:hep-th/0205272].

\bibitem{Karanikas:2003um} A.~I.~Karanikas and C.~N.~Ktorides,
Phys.\ Lett.\ B {\bf 560}, 252 (2003) [arXiv:hep-th/0301214].

\bibitem{Bern:2000mq} Z.~Bern, L.~Dixon and D.~A.~Kosower,
JHEP01 \ (2000)027.

\end{thebibliography}
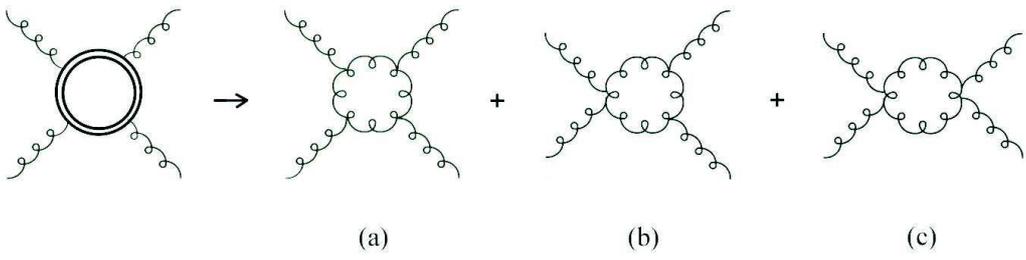
\end{document}